\documentclass[12pt]{elsart}
\usepackage{psfrag}
\usepackage{epsfig}

\newcommand{\beq}{\begin{equation}}
\newcommand{\eeq}{\end{equation}}
\newcommand{\bea}{\begin{eqnarray}}
\newcommand{\eea}{\end{eqnarray}}

\begin{document}

\begin{frontmatter}

\title{Cellular automata for the spreading of technologies in socio-economic systems}

\author{Ferenc Kun},\ead{feri@dtp.atomki.hu} Gergely Kocsis, and J\'anos Farkas
\address{Department of Theoretical Physics\\University of 
Debrecen\\ H-4010 Debrecen, P.O.Box: 5, Hungary}


\begin{abstract}
We introduce an agent-based model for the spreading of technological
developments in socio-economic systems where the technology is mainly
used for the collaboration/interaction of agents. 
Agents use products of different technologies to collaborate with each
other which induce costs proportional to the difference of technological
levels. Additional costs arise when technologies of different
providers are used. Agents can adopt technologies and providers of
their interacting partners in order to reduce their costs leading to
microscopic rearrangements of the system.
Analytical calculations and computer simulations revealed that
starting from a random configuration of different technological levels
a complex time evolution emerges where the spreading of advanced
technologies and the overall technological progress of the system
are determined by the amount of advantages more advanced technologies
provide, and by the structure of the social environment of agents. We
show that agents tend to form clusters of identical technological
level with a power law size distribution. When technological progress
arises, the spreading of technologies in the system can be described by
extreme order statistics. 

\end{abstract}

\end{frontmatter}

\section {Introduction}
Recently, the application of statistical physics and of the theory of
critical phenomena provided novel insight into the dynamics of
socio-economic systems
\cite{weidlich,sznajd_1,sznajd_2,stauffer_1,stauffer_2,stauffer_3,sznajd_3,socio_barcel_1}.
Various types of models have been developed 
which capture important aspects of the emergence of communities
\cite{weidlich}, opinion spreading
\cite{sznajd_1,sznajd_2,stauffer_1,stauffer_2,stauffer_3,sznajd_3} or
the evolution of financial data \cite{bouchaud}. The dynamics of
innovation and 
the spreading of new technological achievements show also interesting
analogies to complex physical systems
\cite{hegselman,silverberg,ruiz,socio_barcel_1}.  
The process of innovation has recently been studied by introducing a
technology space based on percolation theory \cite{silverberg}. In
this model new inventions arise as a result of a random search in the
technology space starting from the current best-practice
frontier. The model could 
reproduce the interesting observation that innovations occur in
clusters whose sizes are described by the Pareto distribution
\cite{silverberg}. Another important aspect of technological
development is the spreading of new technological achievements.
In a socio-economic system
different level technologies may coexist and compete as a result of
which certain technologies proliferate while others disappear from
the system. One of the key components of the spreading of successful
technologies is the copying, {\it i.e.} members of the system adopt
technologies used by other individuals according to certain decision
mechanisms. Decision making is usually based on a cost-benefit balance
so that a technology gets adopted by a large number of individuals if
the upgrading provides enough benefits. The gradual adaptation of high
level technologies leads to spreading of technologies and an overall
technological progress of the socio-economic system. 

In the present paper we consider a simple agent-based model of the
spreading of technological achievements in socio-economic
systems. Agents of the model may represent individuals or firms which
use certain 
technologies to collaborate with each other. For simplicity, we assume
that costs of the cooperation arise solely due to the incompatibility
of technologies used by the agents which then have two origins: on the
one hand, difference of technological levels incurs cost, the larger
the difference is, the higher the cost gets. On the other hand,
technologies used by agents may belong to different providers which
induce additional costs. Agents interacting with their social
neighborhood can decrease their cost by adopting technologies of their
interacting partners. The local rejection-adaptation strategy of
agents can lead to interesting changes of the system on the meso- and
macro-level, namely, agents can form clusters with identical
technological levels, which can also be accompanied by an overall
technological progress of the system. 

We analyze the time evolution of this model
socio-economic system starting from a random configuration of
technological levels and providers without considering the possibility of
innovation. Based on analytic calculations and computer simulations we
study how the adaptation of technologies of interacting partners leads
to spreading of technological achievements. We characterize the
microstructure of communities of agents, and the technological
progress of the system on the macro level. 

\section {Model}
Our model captures some relevant features
of the spreading of technological developments when they are mostly used
for the cooperation of individuals. In the model we
represent the socio-economic system by 
a set of agents which posses products of different technological
levels and use it to cooperate with each other. Thinking in terms of
telecommunication technologies, agents are characterized
by two variables: the technological level of the product an agent has
(the technological level of the device the agent uses for
communication) is described by a real variable $\tau$ such that a
larger value of $\tau$ stands for
more advanced technologies. New technologies developed by the producers
reach the agents through providers. For simplicity, we assume that
there are at most two providers active in the system and each agent
belongs to one of them. The provider of agents is characterized by an
integer variable $S$ which can take two different values $S =
\{-1,1\}$. 

The agents are assumed to cooperate with their social
partners which is the easiest if the partners have products of the same
technological level. Using technologies of different level
can induce difficulties which may be realized by additional costs.
It is reasonable to assume that the cost $C$ induced by the
collaboration of agents $i$ and $j$ is a monotonous function of the
difference 
of the technological levels $|\tau_i-\tau_j|$. For the purpose of the
explicit mathematical analysis we consider the simplest functional
form and cast the cost of cooperation into the following form
\begin{eqnarray}
C(i\to j) = a|\tau_i-\tau_j| + \frac{1}{2}\Delta (1-S_iS_j).
\label{eq:c_simple}
\end{eqnarray}
The equation expresses that being at different technological levels
(having different $\tau$ values) incurs cost, the higher the
difference is in $\tau$ the higher the costs are, while being at the same
technological level is cost-free. This crude assumption models a
socio-economic system which favors the local communities to be at
the same technological level. The value of the multiplication factor
$a$ has to be chosen to capture the effect that in case of different
technological levels it is favorable for 
agents to be on a higher technological 
level than their interacting partners. It follows that the value of
$a$ should depend on the relative technological level of the agents
with the condition  
\begin{eqnarray}
a = 
\left\{
\begin{array}{lll}
 a_1, & if & \tau_i > \tau_j  \\
 a_2, & if & \tau_i < \tau_j  
\end{array} \ \ \ \ \ \ \mbox{where} \ \ \ \ \ a_1 < a_2.
\right.
\label{eq:condition}
\end{eqnarray}
The condition $a_1 < a_2$ implies that being on a higher technological
level, {\it i.e.} being more advanced than the surroundings $\tau_i >
\tau_j$, can lower the costs compared to the opposite case. The second
term of Eq.\ (\ref{eq:c_simple}) takes into account that the cooperation
of agents belonging to different providers implies additional
expenses. We assume that this cost does not depend on the
technological levels by setting $\Delta >0$ to a constant value. Note that
$(1/2)(1-S_iS_j)$ takes value 1 or 0 for agents of different
and of the same providers, respectively, resulting in an additional
cost $\Delta$ when agents of different providers collaborate. The
arrow $\to$ in the argument of $C$ in Eq.\ (\ref{eq:c_simple})
expresses that due to the condition Eq.\ (\ref{eq:condition}) the cost
function is not symmetric with respect to agents $i$ and $j$. Hence,
$C(i \to j)$ defines the cost of agent $i$ arising due to the
cooperation with agent $j$ and this is not equal to the cost of agent
$j$, {\it i.e.} $C(i \to j) \neq C(j \to i)$.

If agent $i$ has $n$ collaborating partners with technological levels
$\tau_1, \tau_2, \ldots , \tau_n$, the total cost of its 
collaboration can be obtained by summing up the cost function Eq.\
(\ref{eq:c_simple}) over 
all connections %taking into account the condition Eq.\ (\ref{eq:condition})
\begin{eqnarray}
C(i) = \sum_{j=1}^{n} C(i \to j).
\label{eq:c_tot}
\end{eqnarray}

\subsection{Dynamics of the model system}
The system agents can change their technological levels with the
aim to minimize their total costs $C(i)$ under the conditions set
by their social environment. 
For simplicity, in the present form of the model no investment is
considered, which means that new technologies cannot appear in the
system. Agents reject their low level technologies and
copy/adopt the more advanced ones of other agents in order to minimize
their costs. We call this microscopic mechanism {\it
rejection-adaptation} which may also improve the global technological
level of the system. Upgrading the technological level of agents does
not induce cost, {\it i.e.} there is no resistance against the change;
if the adaptation is advantageous it will be performed.
The time evolution of the system
proceeds as follows: at time $t$ the communication of agent $i$ with
technological level $\tau_i^t$ and provider $S_i^t$ incurs
the total cost $C^t(i)$. At time $t+1$ the agent can either keep its
technological level or can take over the $\tau$ and $S$ values of one of its
social partners, $\tau_{i}^{t+1} \in \{\tau_i^t, \tau_{1}^t,
\tau_{2}^t, \ldots , \tau_{n}^t \}$ and $S_i^{t+1}\in \{S_i^t, S_1^t,
S_2^t, \ldots , S_n^t\}$. The possibility
which is actually realized is the choice that minimizes the total cost
\bea
C^{t+1}(i) = \min\{C(\tau\in \{\tau_i^t, \tau_{1}^t,
\tau_{2}^t, \ldots , \tau_{n}^t \}, S\in \{S_i^t, S_1^t,
S_2^t, \ldots , S_n^t\}) \}.
\label{eq:dyn_rule}
\eea
Based on this dynamics the time evolution can be followed
by computer simulation treating the system as a cellular automaton,
{\it i.e.}, the dynamics Eq.\ (\ref{eq:dyn_rule}) is evaluated for
each agent at the same time (parallel dynamics) under the neighborhood
conditions set by the structure of the socio-economic environment.

The above dynamics results in a rather complex time evolution of the
system during which certain technologies disappear while others
survive and spread over the system. In the following we analyze the
time evolution of the system on the micro and macro level by varying the
parameters $a_1$, $a_2$, $\Delta$ and the topology of the social
environment of agents. 

\section{Mean field versus local interaction}
\label{sec:mean_field}
In order to understand the decision mechanism how agents select the
technology to adopt, it is useful to study simplified configurations
by analytic calculations. For clarity, first we consider only one
provider in the system, or analogously to set $\Delta =0$.

Let us assume that the system is composed of a large number of agents
which have randomly distributed technological levels in an interval $\tau_{min} \leq
\tau \leq \tau_{max}$ with a probability density $p(\tau)$ and
distribution function $P(\tau) = \int\limits_{\tau_{min}}^{\tau}
p(\tau') d\tau'$. 
In the limiting case of an infinite range of interaction,
all agents interact with all others so that the cost of
interaction of an agent of technological level $\tau$ can be cast into
the form
\bea
\label{eq:mean_field_cost}
C(\tau) = a_1 \int_{\tau_{min}}^{\tau} (\tau - \tau')p(\tau') d\tau' +
a_2 \int^{\tau_{max}}_{\tau} (\tau' - \tau)p(\tau') d\tau'
\eea
as a function of $\tau$. In the next time step the agent will change
its technological level from $\tau$ to that $\tau^*$, which
minimizes the cost function Eq.\ (\ref{eq:mean_field_cost}), {\it
i.e.} $\frac{dc}{d\tau}\left|_{\tau^{*}}\right. = 0$. The technology
that optimizes the cost 
can finally be obtained as the solution of the equation 
\bea
\label{eq:final}
 P(\tau^*) = \frac{1}{1+\frac{1}{r}},
\eea 
where $r=a_2/a_1$ is the ratio of the two cost factors $a_1$ and
$a_2$.
Due to the infinite range of interaction all agents make the same
decision, thus after a single time step all agents adopt the
same technology $\tau^*$ and the evolution of the system stops. It can
be seen from Eq.\ (\ref{eq:final}) that the optimal technology adopted
by the entire system is just determined by the ratio $r = a_2/a_1$
which characterizes how much advantages the more advanced technology
provides compared to the less advanced ones. In the special case of
$r=1$ ({\it i.e.} being on a higher technological level does not
provide any advantages), the system adopts the median $\tau^* = m$ of the
initial distribution of technologies $p(\tau)$ \cite{sornette_1}. It
is interesting to 
note that the optimal choice $\tau^*(r)$ is a monotonically
increasing function of $r$; however (and surprisingly), the most advanced
technology $\tau_{max}$ is solely chosen in the limiting case 
$\lim\limits_{r\to\infty}\tau^*(r) = \tau_{max}$.
At any finite value of $r>1$ the large number of agents of low level
technologies can force the system to stay at a lower technological
level. 

In the following let us consider a finite community of $n$ agents with
technological levels $\tau_1 < \tau_2 < \ldots < \tau_n$ communicating
with each other. The collaboration of agent $i$ of technological level
$\tau_i$ with the other $n-1$ agents induce the cost

\bea
C(\tau_i) = a_1 \sum\limits_{j=1}^{i-1} (\tau_i - \tau_j) + a_2
\sum\limits_{j=i+1}^{n} (\tau_j - \tau_i). 
\label{eq:cost_n}
\eea
In the next time step the agent decides to adopt that technology among
the $n-1$ possibilities which minimizes the cost function Eq.\
(\ref{eq:cost_n}). It can be obtained analytically that the decision
is only determined by $r$, namely, the $i$th largest technological
level is adopted $\tau^* = \tau_i$ when $r$ falls in the interval
\bea
 \frac{i-1}{n-i+1} &<& r < \frac{i}{n-i} \ \ \ \ \mbox{for} \ \ \ 1 \label{eq:interval_1}
\leq i < n, \\ [4mm]
 n - 1 &<& r \ \ \ \ \ \ \ \ \ \ \ \ \ \  \ \mbox{for} \ \ \ i = n.
\label{eq:interval}
\eea
It can be seen from the above equations that the limits of the
subintervals of $r$ to choose the $i$th and $n-(i-1)$th largest $\tau$ are
symmetric with respect to $r=1$.  The most advanced technology $\tau^*
= \tau_{n}$ of the available ones is adopted only if $r$ exceeds the
number of interacting partners $r> n-1$.
Of course, the actual value of $\tau^*$ is not determined by the above
equations, so that in a system composed of 
a large number of local communities of agents with randomly
distributed $\tau$ values a complex time evolution emerges, which is
locally governed by Eqs.\ (\ref{eq:interval_1},\ref{eq:interval}).

\section{Agents on a square lattice}
In order to reveal the time evolution of an ensemble of a large number
of interacting agents based
on the rejection-adaptation mechanism of technologies,
we consider a set of agents organized on a square lattice
of size $L\times L$ with nearest-neighbor interactions. Initially
agents have randomly distributed 
technological levels between 0 and 1 with a uniform distribution
\bea
p_0(\tau) = 1, \ \ \ \ \mbox{and}\ \ \ \  P_0(\tau) = \tau, \ \ \
\mbox{for}  \ \ \ \ 0\leq \tau \leq 1. \label{eq:uniform}
\eea
Assuming periodic boundary conditions, all agents have four
interacting partners. The rejection-adaptation dynamics
based on the cost minimization Eq.\ (\ref{eq:dyn_rule}) results in a
non-trivial time evolution of the system which is followed by computer
simulations treating the system locally as a cellular automaton. It
has to be emphasized that in the simulations parallel update is used,
{\it i.e.} the dynamic rule Eq.\ (\ref{eq:dyn_rule}) is
simultaneously applied to all agents keeping their interacting
partners fixed. This parallel dynamics is one of the sources of the
complex time evolution of the system.

\begin{figure}[!h]
\label{fig:evolv_1p0}
\psfrag{aa}{\large a}
\psfrag{bb}{\large b}
\psfrag{cc}{\large c}
\psfrag{dd}{\large d}
\psfrag{ee}{\large e}
\psfrag{ff}{\large f}
\psfrag{gg}{\large g}
\psfrag{hh}{\large h}
\psfrag{ii}{\large i}
\psfrag{jj}{\large j}
\psfrag{kk}{\large k}
\psfrag{ll}{\large l}
\begin{center}
\epsfig{bbllx=0,bblly=0,bburx=785,bbury=528,file=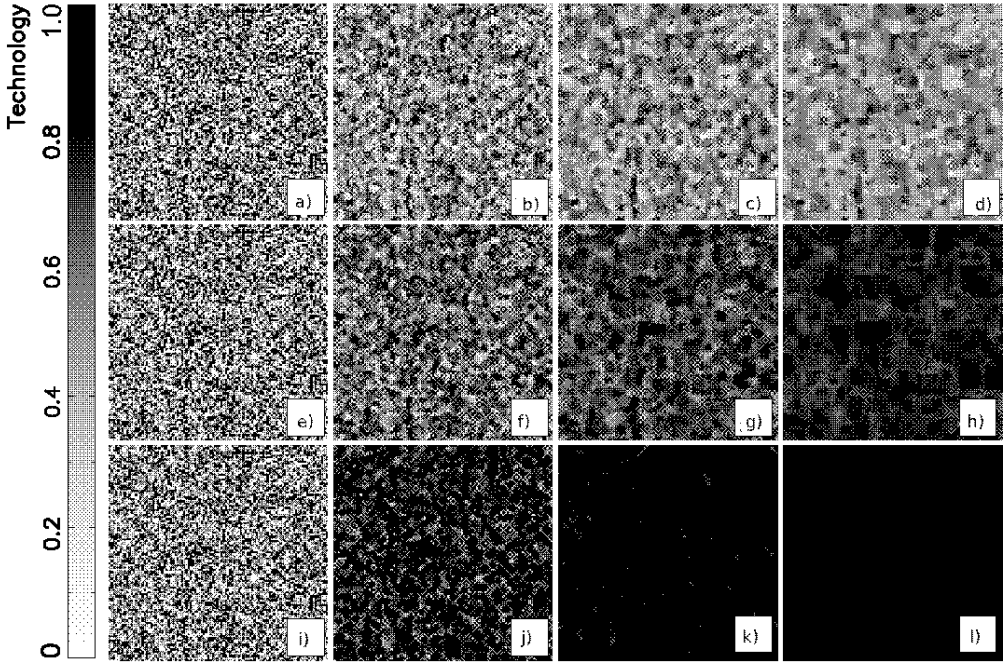,
  width=13.5cm}
 \caption{ Evolution of cellular automata at different
values of $r=a_2/a_1$: r=1.0, 2.0, 4.0 for the upper, middle and lower
rows, respectively. Snapshots were taken at the same times in the
rows: $t=1, 2, 4, 15$ from left to right. The gray-scale illustrates
the technological level of single agents. 
}
\end{center}
\end{figure}
Applying the analytic results Eqs.\
(\ref{eq:interval_1},\ref{eq:interval}) for the 
specific case of $n=4$, the agents will always copy the 1th, 2nd, 3rd
or 4th largest $\tau$ of their local interacting partners when the
value of the parameter $r$ falls in the intervals $0<r< 1/3$, $1/3 < r< 1$, $1<r<3$, $3 <r$,
respectively. Representative examples of the time evolution of a
system of size $L=100$ are shown in Fig.\ 1 for
different parameter values in the range $r \geq 1$.
(Note that the behavior of the system is
symmetric with respect to $r=1$.)  Since the system dynamics
favors local communities to use the same technology (to be
at the same technological level), the agents 
tend to form clusters with equal $\tau$ at any
value of $r$. For the case of
$r=1.0$, when being more advanced than the surroundings does not
provide any advantages, it can be seen that the system evolves into a
frozen cluster structure.
The technological level  $\tau$ of these clusters covers practically
the entire available range with a non-trivial distribution, {\it i.e.}
communities of low level technologies can survive in the presence of
highly advanced ones (see Fig.\ 1$a,b,c,d$). 
At $r=2.0$, where more advanced technologies are favored by the agents
(locally the second largest $\tau$), the system converges into an almost
completely homogeneous state of a relatively high technological
level (see Fig.\ 1$e,f,g,h$). 
In the simulations, initially clusters of agents with identical $\tau$ grow 
and finally the entire system evolves into a homogeneous state with
all agents adopting the same technology. 
\begin{figure}[!h]
\label{fig:cluster_dist_1}
\begin{center}
\epsfig{bbllx=13,bblly=13,bburx=371,bbury=316,file=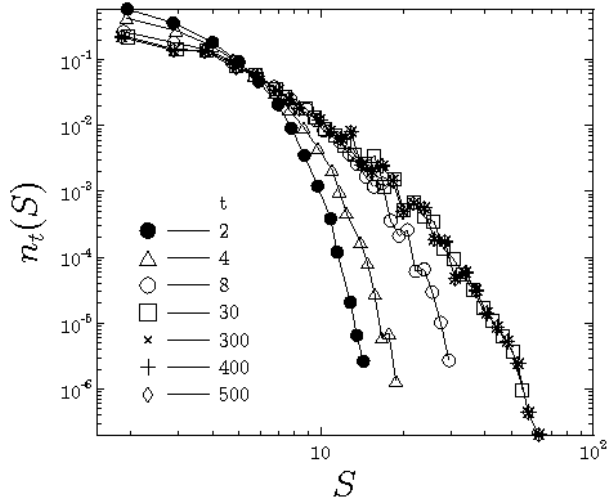,
  width=8.5cm}
\caption{Cluster size distributions $n_t(S)$ at different time values
$t$ for $r=1$. For $t\geq 30$ the distribution practically does not
change, it converges to a rapidly decreasing exponential form.} 
\end{center}
\end{figure}
Locally the agents
decide for the second largest $\tau$, and therefore both very low and
very high level technologies disappear during the evolution.
The gray-scale also illustrates that the limit
value of $\tau$ adopted by almost all agents in Fig.\ 1$h$ falls between 0.8 
and 1, {\it i.e.} it is smaller than the highest available value
$\tau_{max} =1$. 
To reach the most advanced technologies, $r$ has to
surpass the threshold value $r=3$. This regime is illustrated in Fig.\
1$i,j,k,l$ for the specific case of $r=4$, where the
white color in Fig.\ 1$l$ implies that the most
advanced technology $\tau_{max}=1$ spraw onto the entire lattice.

\begin{figure}[!h]
\label{fig:cluster_dist_2}
\psfrag{aa}{\large a)}
\psfrag{bb}{\large b)}
\begin{center}
\epsfig{bbllx=16,bblly=16,bburx=850,bbury=360,file=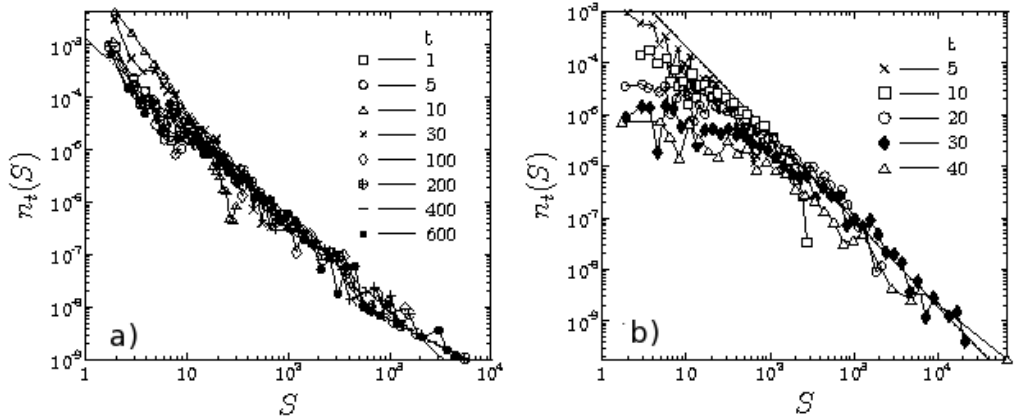,
  width=14.2cm}
\caption{ Cluster size distributions $n_t(S)$ at different time
values $t$ for $r=2$ $(a)$, and $r=4$ $(b)$. 
At the parameter value $r=2.0$ $(a)$ the
distribution $n_t(S)$ evolves from an initially rapidly decreasing exponential
form to a power law distribution. The power law form proved to be the
limit distribution of the system over a broad range of cluster
sizes. The value of the exponent $\alpha$ of the fitted power law (straight
line in the figure) $\alpha = 1.75 \pm 0.05$. In both cases $(a)$ and $(b)$ only one
cluster remains in the final state, 
{\it i.e.} all the agents have the same technological level.
This state is reached through a complex evolution of
cluster sizes. 
} 
\end{center}
\end{figure}
This microscopic restructuring and clusterization process of agents
can be characterized by studying the distribution $n_t$ of cluster
sizes $S$ at different times $t$. A cluster is identified on the
lattice as a connected set of agents with the same technological
level, where the number of agents defines the cluster size $S$. 
The numerical results are presented in Figs.\ 2 and 3 for a system of
size $L=1500$. It can be seen in Fig.\ 2 that for $r=1$, after a few time steps
the cluster size distribution converges to a rapidly decreasing exponential form,
where even the largest cluster contains a relatively small number of
agents. More interesting is the regime $1<r<3$ where agents locally
always prefer to adopt the second highest technological level. In this
case the initially exponential distribution tends to a power law as
time elapses
\beq
n_t(S) \sim S^{-\alpha},
\eeq
and keeps this functional form over a broad range of time scales (see
Fig.\ 3$a$). The
final homogeneous state is reached when small clusters gradually
disappear and only one large cluster remains, but the power law
distribution remains the same for a long time.
The value of the exponent $\alpha$ was determined
numerically as $\alpha = 1.75 \pm 0.05$. For $r>3$ the convergence to the
homogeneous final state is faster, but even in
this case a power law emerges for intermediate times with the same
exponent as before and gradually
disappears as the system gets dominated by a single cluster (see
Fig.\ 3$b$). 

\section{Extreme order statistics and technological progress}
In the previous section we have shown that our rejection-adaptation
mechanism results in a complex time evolution with local
rearrangements which then leads to an overall system change.
A very interesting aspect of the model is that the disappearance
of certain technologies and proliferation of others may give rise to a
global technological progress.
In order to give a quantitative characterization of this evolution
process, we determined the distribution $p_t(\tau)$ of technological
levels $\tau$ at different times, and the
mean $\left<\tau^t\right>$ and standard deviation $\sigma^t$ of
this distribution
\bea
\left<\tau^t \right> = \frac{1}{N}\sum_{i=1}^{N}\tau_i^t, \ \ \ \ \ \ \qquad \qquad
(\sigma^t)^2 = \left<\left(\tau^t - \left<\tau^t \right>
\right)^2\right>. 
\eea
Fig.\ 4 shows that for $r=1$,
when higher level technologies do not provide advantages for agents, the
distribution $p_t(\tau)$ rapidly attains a Gaussian shape. 
\begin{figure}[!h]
\label{fig:dist_evolv_r1}
\psfrag{aa}{\large a)}
\psfrag{bb}{\large b)}
\psfrag{cc}{\large c)}
\begin{center}
\epsfig{bbllx=15,bblly=15,bburx=390,bbury=380,file=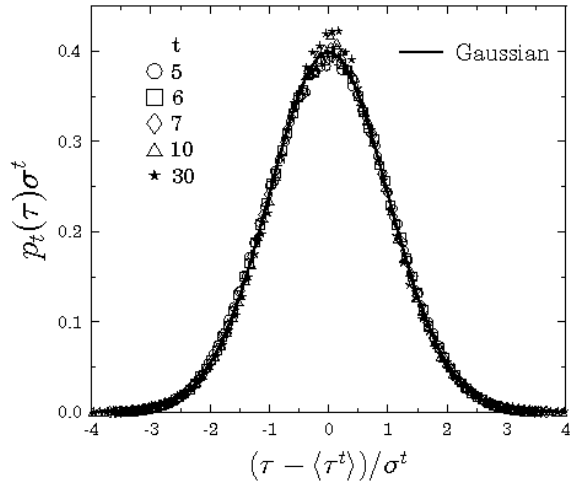,
  width=8.5cm}
\caption{ Rescaled distributions of technological levels for
$r=1$. The standard Gaussian has a perfect agreement with the
numerical results. }
\end{center}
\end{figure}
In order to demonstrate the validity of the Gaussian form, we plot in
Fig.\ 4 the rescaled distributions $p_t(\tau)\sigma^t$ as a function 
of $\left(\tau-\left<\tau^t\right>\right)/\sigma^t$, which
have an excellent agreement with the standard Gaussian
$g(x)=1/\sqrt{2\pi}\exp{(-x^2/2)}$.
The convergence to the Gaussian is very fast, practically
after 30-40 iteration steps the system completely forgets its initial
uniform state and $p_t$ attains the limit distribution.
The Gaussian implies that the fraction of agents having very high and
very low level technologies both decrease and agents tend to copy
technologies in the vicinity of the distribution mean. 
Consequently, the system does not have any technological
progress, the average technological level remains constant during the
time evolution, and  $\left<\tau^t\right> \to 0.5$ (see Fig.\
6$a$). In addition, the standard 
deviation of $p_t$ attains a non-zero constant value in the large $t$
limit, $\sigma^t \to 0.12$ (Fig.\ 6$b$). 

In contrast, when agents locally prefer to adopt higher level
technologies, namely, the largest or the second largest $\tau$ value
of the neighborhood on the square lattice, the system undergoes a more complex
time evolution involving also extreme order statistics. For $1<r<3$
all the agents adopt the second highest available technology; hence,
in a large enough system the distribution of technological levels
right after the first iteration step $p_1(\tau)$ is the $k=3$
rank extreme distribution $\phi^k_M$ of $M=4$ variables of uniform
distribution. 
\begin{figure}[!h]
\label{fig:dist_evolv}
\psfrag{aa}{\large a)}
\psfrag{bb}{\large b)}
\psfrag{cc}{\large c)}
\begin{center}
\epsfig{bbllx=20,bblly=15,bburx=500,bbury=221,file=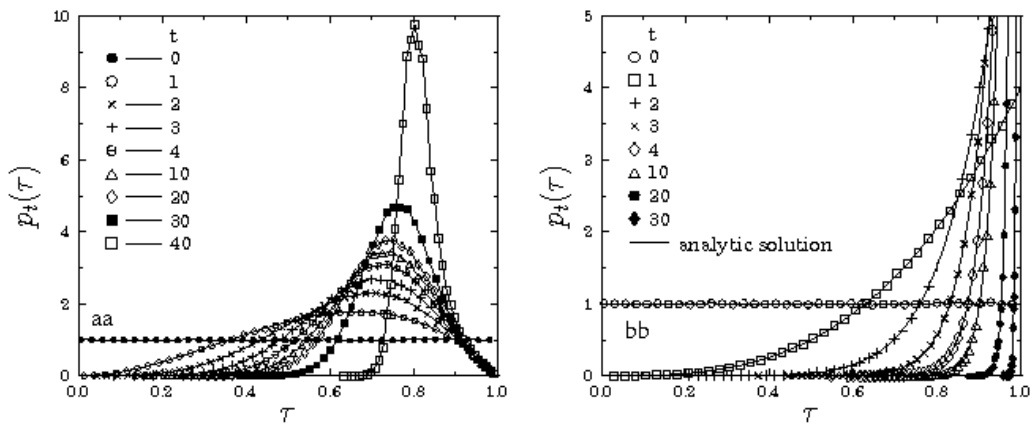,
  width=14.0cm}
\caption{ Distribution $p_t$ of the technological level
$\tau_i^t$ of agents at different time values $t$ during the evolution of
the system. For $r=2$ $(a)$ and $r=4$  $(b)$, the distribution takes an
asymmetric form with an increasing average and decreasing standard
deviation. In $(a)$ the continuous line of $t=1$ indicates $\phi^3_4$,
furthermore, in $(b)$ the lines of $t=1,2,3,4,5$ stand for $\phi^4_4,
\phi^{9}_{9}, \phi^{16}_{16}, \phi^{25}_{25}, \phi^{36}_{36}$, which
are in an excellent agreement with the numerical results.} 
\end{center}
\end{figure}
In general, the probability density function
$\phi^k_M(x)$ of the $k$th largest value of $M$ realizations of the random
variable $x$ which has a probability density $p(x)$ and a distribution
function $P(x)$, can be cast into the form
\beq
\phi^k_M(x) = \frac{M!}{(k-1)!(M-k)!}P(x)^{k-1}(1-P(x))^{M-k}p(x).
\label{eq:extrem}
\eeq
It can be seen in Fig.\ 5$a$ that by substituting the
initial uniform distribution Eq.\ (\ref{eq:uniform}) into Eq.\
(\ref{eq:extrem}), a perfect agreement is obtained 
between $\phi^3_4$ and $p_1(\tau)$. Due to the overlap of the local
neighborhoods of the lattice sites, however, at higher iteration steps
the distributions $p_t$ do not follow Eq.\ (\ref{eq:extrem}) when we
substitute $\phi_M^k$ and the corresponding distribution function
recursively on the right hand side.
By increasing the number of iterations, $p_t$ gets
narrower and converges to a sharply peaked function as the final
homogeneous state is approached (see Fig.\ 5$a$ and
Figs.\ 1$e,f,g,h$). Consequently, the average value
of the technological level increases and converges to a limit value
which is smaller than the available maximum $\tau_{max}=1$. The
standard deviation goes to zero indicating the homogeneity of the
final state (see Fig.\ 6). 
\begin{figure}[!h]
\label{fig:aver_tech}
\psfrag{aa}{\large $a)$}
\psfrag{bb}{\large $b)$}
\begin{center}
\epsfig{bbllx=13,bblly=13,bburx=490,bbury=241,file=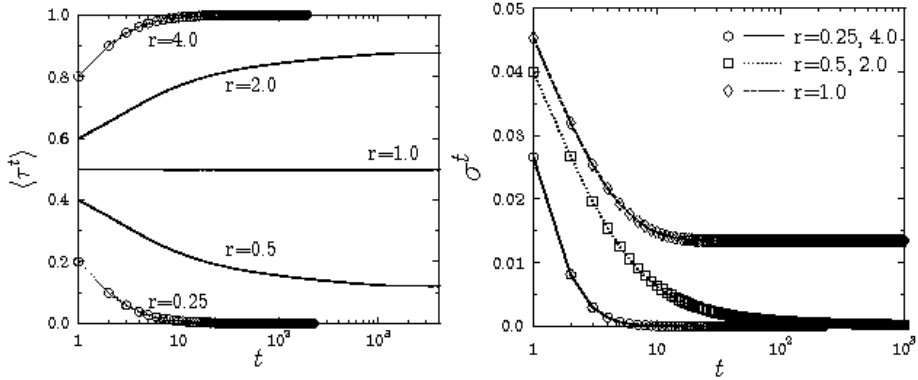,
  width=12.8cm}
\end{center}
\caption{Average technological level $\left< \tau^t\right>$ as a
function of time $t$ $(left)$, and the standard deviation $\sigma^t$
$(right)$. Note the symmetry of the values of $\left< \tau^t\right>$
for $r$ and $1/r$ while the corresponding standard deviations fall on
the top of each other. For $r=4$ and $r=0.25$ the continuous lines in
both figures indicate the analytic results which have an excellent
agreement with the numerical calculations.} 
\end{figure}
When the control parameter $r$ becomes larger than $3$, more advanced
technologies provide so much benefit that it is always better for
agents to adopt the available highest level technology in the local
neighborhood. Consequently, $p_t(\tau)$ rapidly converges to a
sharply peaked form at $\tau_{max}=1$ through extreme order
distributions. It is interesting to note that contrary to the case of
$1<r<3$, in this regime the distribution $p_t$ can be described by the extreme
order density function $\phi^k_M$ Eq.\ (\ref{eq:extrem}) with $k=M$ at any time
$t$ by taking into account that $M$ increases as a function of time. We
found a recursive formula for the time dependence of the parameter $M$
\beq
M_{t+1}=M_t + 5+2(t-1), \qquad \qquad \mbox{with} \qquad \qquad M_1=4, 
\label{eq:iteral}
\eeq
which describes the spreading of successful technologies as a function
of time. The continuous lines in Fig.\
5$b$ demonstrate the excellent agreement of the
above analytic prediction with the numerically obtained distribution
functions. Note that due to the symmetry of the system with respect to
the parameter value $r=1$, the same holds also for $r<1/3$ with
$\phi^1_M$, where the smallest value ($k=1$) of $M_t$ variables given
by Eq.\ (\ref{eq:iteral}) is selected. 
These results imply that the average technological level in these
regimes can easily be obtained analytically, {\it i.e.} the average
of the largest and of the smallest value of $M_t$ variables with uniform
distribution can be cast into the form 
\beq
\left<\tau_{max}\right>=M_t/(M_t+1), \ \ \ \ \qquad  \qquad
\left<\tau_{min}\right>=1/(M_t+1).  
\label{eq:min_max_aver}
\eeq
Substituting the recursive formula of $M_t$ into Eq.\
(\ref{eq:min_max_aver}) a perfect agreement is obtained with the
numerical results of $\left<\tau^t\right>$ presented in Fig.\
6$a$. 

\section{Two providers}
By now we have studied the behavior of the system without
considering the effect of providers, {\it i.e.} all the agents
belonged to the same provider. In the following we show that
the presence of more than one provider results in a substantial
change of the behavior of the system on the meso- and macro-levels.  
\begin{figure}[!h]
\label{fig:delta}
\begin{minipage}[c]{0.45\textwidth}
\psfrag{aa}{\large $(a)$}
\begin{center}
\epsfig{bbllx=13,bblly=13,bburx=380,bbury=320,file=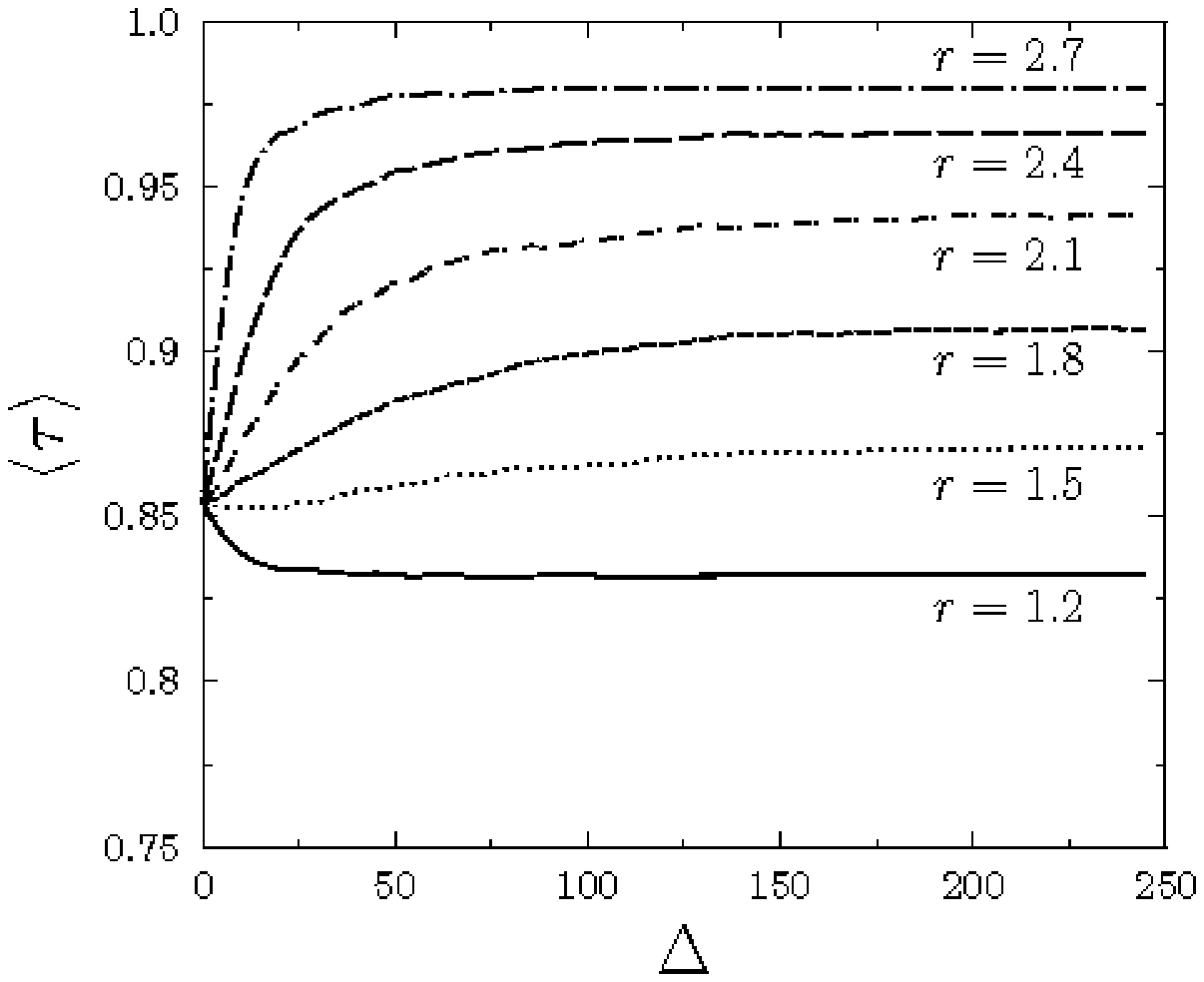,
  width=7.3cm}\hspace*{0.5cm}
\end{center}
\end{minipage}
\begin{minipage}[r]{0.45\textwidth}
\psfrag{bb}{\large $(b)$}
\begin{center}
\hspace*{0.8cm}\epsfig{bbllx=13,bblly=13,bburx=380,bbury=320,file=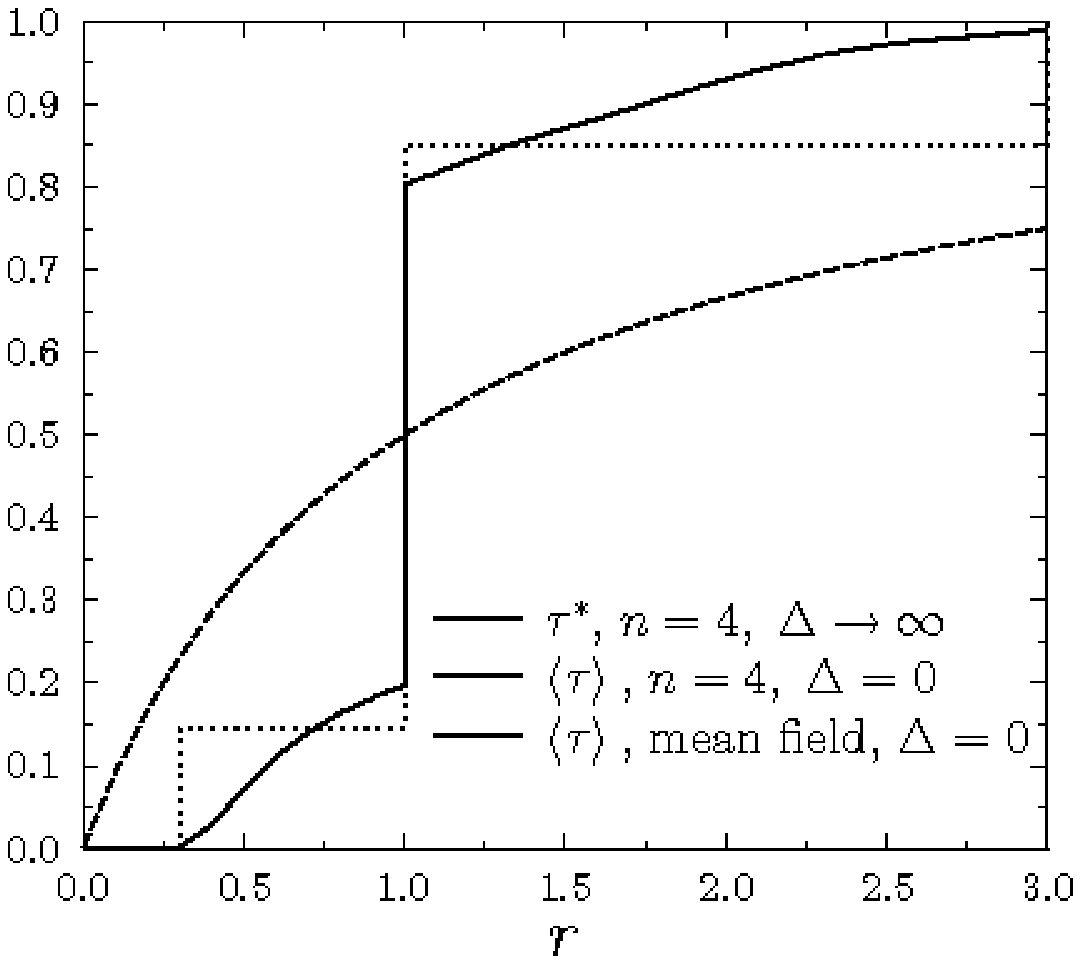,
  width=6.9cm}
\end{center}
\end{minipage}
\caption{$(left)$ The asymptotic value of the average technological level
$\left<\tau\right> = \lim\limits_{t\to\infty}\left<\tau^t\right>$
as a function of $\Delta$. 
For large $\Delta$ the value of $\left<\tau\right>$ converges to a
limit value $\tau^*$, which depends on $r$. $(right)$
$\tau^*$ as a function of $r$ together with the mean
field result Eq.\ (\ref{eq:final}) and the analytic solution Eqs.\
(\ref{eq:interval_1},\ref{eq:interval}) for
$n=4$ obtained at $\Delta=0$.}  
\end{figure}

Let us assume that the initial technological levels are randomly
distributed, as before, but in addition a number $n_1$ of agents have
a provider $S=+1$ and a 
number $n_2$ has another one $S=-1$. The value of $S$ is
independent of the technological levels $\tau$. The fraction of agents
belonging to the two providers are $p=n_1/N$ and $q=1-p=n_2/N$, where
$N=n_1+n_2$ is the total number of agents. In the case of an infinite
range of interaction   
among agents (mean field approach) it is straightforward to show that
the cost function $C(\tau,\Delta)$ at a finite value of $\Delta$ can
be cast into the form  
\beq
C(\tau,\Delta) = C(\tau) + 2pq\Delta,
\eeq
where $C(\tau)$ denotes the mean field cost function Eq.\
(\ref{eq:mean_field_cost}) where no providers were considered.  
Based on the arguments presented in Sec.\ \ref{sec:mean_field}, it
follows that after one iteration step the system minimizes the cost
function by attaining a uniform state where the median $m$ of the initial 
distribution of technological levels is copied by all agents and
they choose the same provider, namely, the one with the higher
initial fraction.

A more interesting (and more realistic) situation occurs when agents
of the two providers separate according to the technological levels,
{\it i.e.} we assume that the $n_1$ agents of the $S=-1$ provider have
the same technological level $\tau_1$, while the $n_2$ agents of the
$S=+1$ provider have a different one $\tau_2$, where $\tau_1 >\tau_2$.
We can determine analytically that even if higher level technologies
provide advantages ($r>1$), the agents choose the lower level technology
$\tau_2$ to minimize 
$C(\tau,\Delta)$ if the excess cost $\Delta$ induced by the
interaction of agents of different providers surpass a threshold value
\beq
\frac{n_2a_2-n_1a_1}{n_1-n_2}(\tau_1-\tau_2) < \Delta.
\label{eq:two_prov_two_tech}
\eeq
%It can be seen that the threshold value of $\Delta$ depends on the
%fraction of agents of the two subsets of providers and on the
%difference of the technological levels $\tau_1$ and $\tau_2$.
%It has to be emphasized that in Eq.\ (\ref{eq:two_prov_two_tech}) an
%explicit dependence on the cost 
%factors $a_1$ and $a_2$ of $C(\tau,\Delta)$ occurs instead of the
%ratio $r=a_2/a_1$. 
In the specific case of $n_1=n_2$ the decision of
agents does not depend on the value of $\Delta$, it is only determined
by the ratio $r$, as it was the case when there was only one
provider present in the system ($n_1 = 0$ or $n_2=0$). In local
communities (smaller groups of agents) in an extended system such
inhomogeneous configurations can frequently occur when two different
providers have different number of agents with a clear separation of
their technological levels. The above results imply that in such cases
$\Delta$ can have a substantial effect on the behavior of the system
by modifying the optimal decision of agents with respect to the
$\Delta=0$ case. In order to give a quantitative characterization of
the effect of $\Delta$ on the micro- and macro-level of the
socio-economic system, we carried out computer simulations on a square
lattice varying the value of $\Delta$ in a broad range for fixed $r$
parameters. The calculations started from a uniform distribution of
technological levels where the two providers had the same fraction
$p=0.5$ and $q=0.5$. Computer simulations revealed that as time
elapses the average technological level $\left<\tau^t\right>$
converges to a limit value which depends both on $r$ and $\Delta$. 
Fig.\ 7$a$ presents representative examples
of the large time limit of $\left< \tau^t\right>$ as a function
of $\Delta$ for several different values of $r=a_2/a_1$ in the range
$1<r<3$. It can be seen that at $\Delta=0$ all the curves obtained at
different $r$ values start from the same point; however, as $\Delta$
increases the curves split up and the system follow $r$-dependent
histories. For very large $\Delta$ the system converges
to a limit value $\tau^*(r)$ which solely depends on the parameter
$r$.  Fig.\ 7$b$ presents a comparison of the $\tau^*$ 
obtained analytically, results of
computer simulations performed on a square lattice with $\Delta=0$
(one provider), and the $\Delta \to \infty$ limit values $\tau^*(r)$ of $\left<
\tau\right>$ as a function of $r$. It is interesting to note that
in the regime $r>1$ for most of the cases the presence of two
providers makes possible a more intense technological progress, {\it
i.e.} agents have a higher tendency to adopt technologies closer to the
possible maximum compared to the case of a single provider. As
$\Delta$ increases the system converges to a steady state where the
average technological level is higher than it was with a single
provider. 

\section{Discussion}

We presented an agent-based cellular automata model to study the
spread of technological achievements in socio-economic systems. Agents
of the model can represent individuals or firms which use different
level technologies to collaborate with each other. Costs arise due to
the incompatibility of technological levels and to different
technological providers. Agents can reduce their costs by adopting
the technologies and providers of their interacting partners. We
showed by analytic calculations and computer simulations that the local
adaptation-rejection mechanism of technologies results in a complex time
evolution accompanied by microscopic rearrangements of technologies
with the possibility of technological progress on the macrolevel. We
showed that agents tend to form clusters of equal technological
levels. If higher level technologies provide advantages for agents,
the system evolves to a homogeneous state but clusters show a power
law size distribution for intermediate 
times. The redistribution of technological levels involves extreme
order statistics leading to an overall technological progress of the
system. The presence of providers proved to play a substantial role in
the time evolution. The competition of providers seems to make the
system more sensitive to advantages provided by the higher level
technologies and can lead to 
additional technological progress by forcing the agents to select locally
the more advanced technology. 

Our model emphasizes the importance of copying in the spreading of
technological achievements and considers one of the simplest possible
dynamical rules for the decision mechanism. In the model calculations
no innovation was considered, {\it i.e.}, 
agents could not improve their technological level by locally
developing a new technology instead of only taking over of the technology
of others. Innovation in the model can be taken into account by
randomly selecting agents to increase their technological level by
a random amount according to some probability distribution. The
generalization of the model in this direction is in
progress. Our calculations show also the importance of the structure
of local communities in the time evolution of the system which
addresses interesting questions for future studies of the model
varying the coordination number of the lattice, and on small-world
and scale-free network topologies
\cite{stauffer_1,stauffer_3,sznajd_3,network,network_1}. The emergence
of power law size distribution of clusters of agents with equal
technological level and the behavior of the exponents on different
topologies can be relevant also for applications.

Compared to opinion spreading models like the Sznajd-model
\cite{sznajd_1,sznajd_2} and its variants
\cite{stauffer_1,stauffer_3,sznajd_3},
the main difference is that in our case the technological level of
agents is a continuous random variable; furthermore, the decision
making is not a simple majority rule but involves a minimization
procedure. A closer analogy can be found when two providers are
considered in the system so that the spreading of a provider could be
interpreted as a success of one of two competing ``opinions''. Opinion
of individuals can also be represented by a continuous real variable
which makes possible to study under which conditions consensus,
polarization or fragmentation of the system can occur
\cite{hegselman}. Such models show more similarities to our spreading
model of technologies.

It is interesting to note that our model captures some of the key
aspects of the spreading of telecommunication technologies, where for
instance mobile phones of different technological levels are used by
agents to communicate/interact with each other. In this case, for
example, 
the incompatibility of MMS-capable mobile phones with the
older SMS ones may motivate the owner to reject
or adopt the dominating technology in his social neighborhood by taking
into account the offers of providers of the interacting partners.

\section*{Acknowledgment:}
This work was carried out with the generous support of Toyota Central
R\&D Labs., Aichi, Japan. F.\ Kun was also supported by OTKA
T049209. We would like to thank our referees for the valuable
comments and suggestions.


\begin{thebibliography}{99}

\bibitem{weidlich}W.\ Weidlich, Sociodynamics: A systematic approach to
mathematical modelling in the social sciences, (Dover Publications,
Mineola, USA, 2000).
\bibitem{sznajd_1}K.\ Sznajd-Weron and J.\ Sznajd, Int.\ J.\ Mod.\
Phys.\ C {\bf 11}, 1157 (2000). 

\bibitem{sznajd_2}K.\ Sznajd-Weron and R.\ Weron, Int.\ J.\ Mod.\
Phys.\ C {\bf 13}, 115 (2002). 

\bibitem{stauffer_1}D.\ Stauffer, Journal of Artificial Societies and Social
Simulation {\bf 5}, No.\ 1, paper 4. 
\bibitem{stauffer_2}D.\ Stauffer, Int.\ J.\ Mod.\ Phys.\ C {\bf 13},
315 (2002). 
\bibitem{stauffer_3}A.\ T.\ Bernardes, D.\ Stauffer, and J.\
Kert\'esz, Eur.\ Phys.\ J.\ B {\bf 25}, 123 (2002).
\bibitem{sznajd_3}K.\ Sznajd-Weron, Acta Physica Polonica B {\bf 36},
2537 (2005).
\bibitem{hegselman}R.\ Hegselmann and U.\ Krause, Journal of
Artificial Societies and Social Simulations {\bf 5}, No.\ 3, paper 2.
\bibitem{silverberg}G.\ Silverberg and B.\ Verspagen, Journal of
Economic Dynamics and Control {\bf 29}, 225 (2005).
\bibitem{ruiz}R.\ M.\ Ruiz, E.\ Albuquerque, L.\ C.\ Ribeiro, and A.\
T.\ Bernardes, AIP Conf.\ Proc.\ {\bf 779}, 162 (2005).
\bibitem{socio_barcel_1}
A.\ Arenas, A.\ D\'\i az-Guilera, C.\ J.\ P\'erez, and F.\ Vega-Redondo,
Phys.\ Rev.\ E {\bf 61}, 3466 (2000).
\bibitem{bouchaud}J.-P.\ Bouchaud and M.\ Potters, {\it Theory of
financial risk and derivative pricing}, (Cambridge University Press,
2000). 

\bibitem{sornette_1}
D.\ Sornette, {\it Critical Phenomena in Natural sciences}, (Springer
Verlag, Berlin, 2000).

\bibitem{network}A.\ L.\ Barabasi and R.\ Albert, Science {\bf 286},
509 (1999). 
\bibitem{network_1} M.\ E.\ J.\ Newman, SIAM Review {\bf 45}, 167 (2003). 
\end{thebibliography}
\end{document}